\documentclass[conference]{IEEEtran}
\IEEEoverridecommandlockouts
\usepackage{cite}
\usepackage{balance}
\usepackage{amsmath,amssymb,amsfonts}
\usepackage{algorithmic}
\usepackage{graphicx}
\usepackage{textcomp}
\usepackage{xcolor}
\usepackage{tikz}
\usepackage{booktabs}
\usepackage{amssymb}
\usepackage{pifont}
\newcommand{\cmark}{\ding{51}}%
\newcommand{\xmark}{\ding{55}}%

\def\BibTeX{{\rm B\kern-.05em{\sc i\kern-.025em b}\kern-.08em
    T\kern-.1667em\lower.7ex\hbox{E}\kern-.125emX}}
\begin{document}

\title{Outdoor Environment Reconstruction with Deep Learning on Radio Propagation Paths}

\author{
\IEEEauthorblockN{ 
Hrant Khachatrian\IEEEauthorrefmark{1}\IEEEauthorrefmark{2}, 
Rafayel Mkrtchyan\IEEEauthorrefmark{1}\IEEEauthorrefmark{2}, 
Theofanis P. Raptis\IEEEauthorrefmark{3}
}
\IEEEauthorblockA{
\IEEEauthorblockA{\IEEEauthorrefmark{1}Yerevan State University, Yerevan, Armenia. Email: \{hrant.khachatrian, rafayel.mkrtchyan\}@ysu.am}
\IEEEauthorblockA{\IEEEauthorrefmark{2}YerevaNN, Yerevan, Armenia}
\IEEEauthorrefmark{3}Institute of Informatics and Telematics, National Research Council, Pisa, Italy. Email: theofanis.raptis@iit.cnr.it
}
}

\maketitle

\begin{tikzpicture}[remember picture,overlay]
\node[anchor=south,yshift=10pt] at (current page.south) {\fbox{\parbox{\dimexpr\textwidth-\fboxsep-\fboxrule\relax}{
  \footnotesize{
    This work has been submitted to the IEEE for possible publication. Copyright may be transferred without notice, after which this version may no longer be accessible.
  }
}}};
\end{tikzpicture}

\vspace{-0.5cm}

\begin{abstract}
Conventional methods for outdoor environment reconstruction rely predominantly on vision-based techniques like photogrammetry and LiDAR, facing limitations such as constrained coverage, susceptibility to environmental conditions, and high computational and energy demands. These challenges are particularly pronounced in applications like augmented reality navigation, especially when integrated with wearable devices featuring constrained computational resources and energy budgets. In response, this paper proposes a novel approach harnessing ambient wireless signals for outdoor environment reconstruction. By analyzing radio frequency (RF) data, the paper aims to deduce the environmental characteristics and digitally reconstruct the outdoor surroundings. Investigating the efficacy of selected deep learning (DL) techniques on the synthetic RF dataset WAIR-D, the study endeavors to address the research gap in this domain. Two DL-driven approaches are evaluated (convolutional U-Net and CLIP+ based on vision transformers), with performance assessed using metrics like intersection-over-union (IoU), Hausdorff distance, and Chamfer distance. The results demonstrate promising performance of the RF-based reconstruction method, paving the way towards lightweight and scalable reconstruction solutions.
\end{abstract}


\section{Introduction}

Outdoor environment reconstruction refers to the process of creating digital representations of outdoor spaces, encompassing various features such as terrain, buildings, and infrastructure. This task plays a crucial role in urban planning \cite{YI20171}, augmented reality \cite{10.1145/2047196.2047270}, simultaneous localization and mapping \cite{10227894} and industrial network optimization \cite{9039732}. Accurate reconstruction of outdoor environments facilitates tasks such as infrastructure deployment planning, signal propagation analysis, and resource allocation.

Traditionally, outdoor environment reconstruction has relied heavily on vision-based methods, utilizing techniques such as photogrammetry and Light Detection and Ranging (LiDAR). While these methods can provide detailed visual representations of the environment, they often face challenges such as limited coverage, susceptibility to environmental conditions (e.g., weather, lighting), and high computational costs. Moreover, vision-based approaches may struggle to capture certain features accurately, especially in densely vegetated or occluded areas \cite{JIAN20209554}. For example, the effectiveness of structured light cameras and time-of-flight cameras can be significantly obstructed by string sunlight \cite{7177380}, \cite{8785774}, and of binocular vision by weak textures and repeated areas \cite{Fan2016}. 

As an example, in more futuristic application scenaria, such as augmented reality navigation in urban environments, the aforementioned traditional energy-intensive and computation-heavy environment reconstruction methods can pose significant challenges, especially when integrated with pervasive wearable devices, such as smart glasses \cite{9166715}. However, the limited computational resources available on such wearable devices, can make it burdensome or infeasible to process the large volumes of input data required for traditional methods \cite{9556496}. Moreover, the energy consumption associated with continuous data processing can quickly deplete the battery life of wearable devices, limiting their usability and practicality for extended navigation tasks. To address these challenges, a more lightweight implementation can be achieved by introducing the concept of using the ambient radio frequency (RF) signals, as additional input data for environment reconstruction. 

The analysis of RF data presents new opportunities for adding a new dimension on the input data or even completely addressing localization-type kind of problems such as node positioning \cite{our-localization-BDS2023} or outdoor environment reconstruction. RF signals propagate through the environment and interact with various elements such as buildings and terrain. By analyzing RF data, it becomes possible to infer the characteristics of the surrounding environment and even reconstruct it in digital form. Unlike vision-based methods, RF-based approaches are less affected by environmental conditions and can provide valuable insights into the structural and material properties of objects.

However, one of the major challenges in leveraging RF data for outdoor environment reconstruction is the scarcity of large-scale, real-world datasets. Collecting comprehensive RF datasets in diverse outdoor environments is a costly and time-consuming endeavor. To address this issue, researchers have turned to synthetic datasets generated using realistic propagation models. Synthetic datasets offer the advantage of providing labeled data at scale, enabling the training of data-hungry deep learning models without the need for extensive data collection efforts.

Even with access to RF measurements though, the reconstruction task remains challenging due to the inherent properties and weaknesses of wireless signals. RF signals are susceptible to various factors such as multipath propagation, signal attenuation, and interference, which can distort the received signal and introduce uncertainties in the reconstruction process. Moreover, the non-line-of-sight (NLOS) nature of RF propagation in outdoor environments further complicates the task, as signals may reflect off surfaces or diffract around obstacles, leading to complex signal patterns that are difficult to interpret accurately. Addressing these challenges requires robust algorithms capable of effectively modeling and exploiting the complex interactions between RF signals and the environment.

Despite the potential of synthetic RF datasets, there remains a notable gap in research focused on utilizing deep learning techniques for outdoor environment reconstruction using such datasets. While deep learning has demonstrated remarkable success in various domains, its application to synthetic RF datasets for outdoor environment reconstruction remains relatively unexplored. This paper aims to fill this gap by investigating the effectiveness of deep learning approaches on a selected synthetic RF dataset for reconstructing outdoor environments.

\emph{Our contribution}: In this paper, we (a) designed an evaluation framework for outdoor environment reconstruction using only radio signal information, (b) designed convolutional and transformer-based neural networks for environment reconstruction that leverage two distinct representations of radio links, obtained a model with 42.2\% intersection-over-union on the WAIR-D dataset, (c) examined strengths and weaknesses of the proposed models, identified prediction errors that are not recoverable based on purely radio link information.

\section{Related works}


In recent years, a diverse range of technologies has been used for achieving environment reconstruction, addressing various challenges and opportunities in urban planning, agriculture, engineering, and robotics domains. One notable study focuses on the digital transformation of urban planning processes, emphasizing the use of 3D spatial data and models to create a city digital twin, enabling more illustrative and comprehensible representations of urban environments \cite{schrotter_digital_2020}. In the agricultural robotics domain, another study proposes a virtual reality (VR) and Kinect-based immersive teleoperation system for navigating unstructured agricultural environments, utilizing real-time 3D reconstruction algorithms to create realistic virtual environments \cite{CHEN2020105579}. Additionally, digital twin technology has gained traction in engineering communities, with a study presenting an AI-powered framework for efficient communication and reconstruction of large-scale digital twins using 3D point cloud data \cite{10242296}. Furthermore, a decentralized framework is proposed for collaborative 3D mapping of outdoor areas using mobile ground robots, demonstrating the reliability and efficiency of real-time 3D LiDAR measurements and peer-to-peer communication strategies \cite{s23010375}. A more resource-heavyweight approach presented in another study leverages signal-to-noise ratio (SNR) measurements from low earth orbit (LEO) communication satellites for real-time 3D city map reconstruction, offering a novel solution to overcome the limitations of traditional passive sensors and enable global-scale mapping \cite{9880764}. However, despite these advancements, challenges persist in achieving cost-effective, lightweight and scalable 3D map reconstruction for urban environments. 

Our paper represents an innovative effort in the field of environment reconstruction by exclusively leveraging RF signals. To the best of our knowledge, this is the first attempt of its kind, breaking new ground in the field of spatial mapping. The initial conceptualization of such idea was introduced in the WAIR-D dataset \cite{huangfu2022waird}, which serves as a foundational reference for our work. In our experiments, where we in fact utilize the WAIR-D \cite{huangfu2022waird}, we take this a step further via exploring the feasibility and efficacy of reconstructing environments solely through RF signals, thereby advancing the frontier of environment reconstruction methodologies.

\section{Problem formal definition}\label{sec::problem}
In this section we follow the notation used in \cite{our-localization-BDS2023}.

Let $u_l$ denote the $l$-th urban area, also referred to as a map. It is represented with a single-channel binary image $\{0,1\}^{w \cdot h}$ where $w$ and $h$ are the width and height of the map, pixels having value 1 correspond to buildings, and 0-valued pixels correspond to free areas including roads. Each map contains a set $V_l$ of user equipments (UEs) and another set $W_l$ of base stations or antennas (BSs). We are given locations of UEs $v_i \in \mathbb{R}^2$ and BSs $w_j \in \mathbb{R}^2$, and radio propagation parameters between every pair. For every UE $i \in V_l$ and BS $j \in W_l$, we have a \textit{radio link} consisting of $K_{i,j}$ distinct radio paths. The $k$-th radio path is described by three parameters: angle of arrival (AoA) at the BS denoted by $\psi_{i,j,k}$, angle of departure (AoD) from the UE denoted by $\phi_{i,j,k}$, and path delay, i.e. the time it takes for the signal to travel from the UE to the BS along the $k$-th path, denoted by $\tau_{i,j,k}$. The \textit{descriptor} of the $k$-th path is the tuple $p_{i,j,k} = \left( \psi_{i,j,k},\phi_{i,j,k},\tau_{i,j,k} \right)$. We also define $R_{i,j} = \{p_{i,j,k} : k=1,\ldots,K_{i,j}\}$ as the set of all paths between the given pair.

Informally, the goal is to reconstruct the map given the locations of the UEs and BSs along with the descriptors of the radio paths between them. Formally, we are looking for an estimator $f\left(\{v_i\}_{i \in V_l}, \{w_j\}_{j \in W_l}, \{R_{i,j}\}_{i \in V_l, j \in W_l}\right)$ that outputs a binary matrix $\hat{u}_l \in \{0,1\}^{w \cdot h}$, called \textit{predicted map}, and minimizes the following objective:
\begin{align*}
    \frac{1}{L}\sum_{l=1}^{L}{\text{IoU}\left(u_l, f\left(\{v_i\}_{i \in V_l}, \{w_j\}_{j \in W_l}, \{R_{i,j}\}_{i \in V_l, j \in W_l}\right)\right)}
\end{align*}
Here, the $\text{IoU}(a,b) = \frac{|a \wedge b|}{|a \vee b|}$ denotes intersection-over-union \cite{iou} defined over two binary matrices of equal size, where $|z|$ operator denotes the sum of all elements of the binary matrix $z$.

In addition to IoU, we are going to evaluate predictors with several other metrics. Some metrics interpret each map as a union of polygons without interior, one polygon per building. Let $P_u$ denote the set of points of the polygons of the map $u$. For any point $p \in P_{u_1}$, $d(p, P_{u_2})$ denotes the distance from $p$ to the closest point from $P_{u_2}$. Hausdorff distance is a directed metric that measures the maximum of those distances over all polygons from one map to another, while Chamfer distance measures the average of the same distances. 
\begin{itemize}
    \item Precision, formally defined as $\frac{1}{L}\sum_{l=1}^{L}{\frac{|\hat{u}_l \wedge u_l|}{|\hat{u}_l|}}$
    \item Recall, formally defined as $\frac{1}{L}\sum_{l=1}^{L}{\frac{|\hat{u}_l \wedge u_l|}{|u_l|}}$
    \item Hausdorff distance \cite{hausdorff_grundzuge_1914}, defined as 
    \begin{align*}
        \max\left(\max\limits_{p \in P_{u_l}}{d(p, P_{\hat{u}_l})}, \max\limits_{p \in P_{\hat{u}_l}}{d(p, P_{u_l})}\right)
    \end{align*}
    \item Chamfer distance \cite{10.5555/1622943.1622971}, defined as 
    \begin{align*}
        \frac{1}{2}\left(\frac{1}{|P_{u_l}|}\sum\limits_{p \in P_{u_l}}{d(p, P_{\hat{u}_l})} + \frac{1}{|P_{\hat{u}_l}|}\sum\limits_{p \in P_{\hat{u}_l}}{d(p, P_{u_l})}\right)
    \end{align*}
\end{itemize}
Note that Hausdorff distance can be sensitive to outlier points on the map.

\section{DL-driven approaches} \label{sec::ml}

\subsection{Input representations}
In this section we describe two deep-learning based approaches to solve the problem described above. As the expected output is a binary map, we can easily utilize neural architectures that output maps, i.e. used for image segmentation tasks. On the other hand, the structure of the inputs of the predictor are more complicated and do not fit regular deep learning pipelines. Note that we have $|V_l||W_l|$ BS-UE pairs for each map $u_l$. Each of these pairs are described with two 2D points, $v_i$ and $w_j$, and a set of $K_{i,j}$ paths, each described by a tuple $p_{i,j,k} = \left( \psi_{i,j,k},\phi_{i,j,k},\tau_{i,j,k} \right)$. 

Inspired by \cite{our-localization-BDS2023}, we suggest two distinct ways to encode this information: as a multi-channel image and as a set of feature vectors. Every pair of BS and UE can be encoded as two real-valued matrices of the same size as the map $u_l$ corresponding to AoA and AoD. In the AoA channel, for each radio path we draw a ray starting at the location of the base station $v_i$ and going in the direction of the AoA. We fill the pixel values with the corresponding time of arrival value divided by the size (in meters) of the environment. AoD channel is the same but the rays start from the UE location $w_j$. For each environment we end up with $2|V_l||W_l|$ matrices or channels.

The second representation of each UE-BS pair is a set of $K_{i,j}$ vectors, each of them being a concatenation of $v_i$, $w_j$ and $p_{i,j,k}$.

\subsection{Convolutional approach}
Having the first multi-channel representation of the input we can use the most commonly used fully convolutional neural network initially designed for semantic segmentation, U-Net \cite{unet}. To optimize the number of image channels, we combine all the channels corresponding to the same base station, separately for AoD and AoA, by taking the maximum over all corresponding channels, thus ending up with $2|V_l|$ channels. We then feed this to the U-Net model and expect the reconstructed map at the output. We use dice loss \cite{dice}. In one experiment we tried to combine it with a regular cross entropy loss. To avoid overfitting, in each epoch we randomly subsample $\frac{1}{2}|V_l|$ to $\frac{1}{2}|V_l|$ UEs and $\frac{1}{2}|W_l|$ to $|W_l|$ base stations. The channels corresponding to the filtered out UE or BS were filled with zeros to ensure the number of input channels remains the same.

We tried to incorporate the second representations of the same radio links as well by passing them through multi-layer perceptrons or transformers and then concatenating the results to U-Net's bottleneck layer, but were not able to get any significant improvement. 

For one experiment we decided to initialize the encoder with ImageNet-pretrained weights. We used U-Net implementation from Pytorch Bolts library which does not provide pretrained weights for that particular architecture. Instead, we used pretrained ResNet-50, a similar convolutional encoder. ResNet-50 has two more convolutional block than the regular U-Net encoder, so we had to add additional two blocks in the decoder as well for symmetry. Also, as the pretrained ResNet expects 3 channels at the input, we used a convolutional layer with 1$\times$1 kernel to convert $2|V_l|$ input channels to 3 channels.

\subsection{Transformer-based approach}
To properly incorporate second kind of representations of the radio links in addition to the ray-based maps, we designed a transformer-based architecture. It is based on Vision Transformers (ViTs \cite{vit}) which can accept multi-channel images split into small patches. We designed our network to be compatible with CLIP's \cite{clip} vision transformer so we can initialize from its pretrained weights. We combine AoA and AoD channels for each UE-BS pair, and add an additional convolutional layer with 1$\times$1 kernel to transform $|V_l||W_l|$ input channels into 3 channels. In addition to patch tokens coming from the multi-channel input, we pass additional tokens, one per each UE-BS pair. Each of these tokens is constructed by a multi-layer perceptron (MLP) with two hidden layers of size 256 and 1024, that accepts the raw features of the 5 shortest radio paths. If there are not enough radio paths we fill the corresponding vector with zeros. We call this model CLIP+. 

To avoid overfitting we also randomly sample 10 to $|V_l||W_l|$ UE-BS pairs. Again, the channels corresponding to removed pairs are filled with zeros, and the corresponding tokens are just excluded. Hence, the number of the tokens in the output can vary. To obtain a fixed sized representation, we add a convolutional layer with 1$\times$1 kernel on top of the ViT to map the CLS token, image patch output embeddings and the first ten radio link token embeddings generated by the CLIP+ model to 196 tokens. Then we reshape the result into 14$\times$14 squares and feed them to UPerNet convolutional module \cite{upernet}, which outputs a 14$\times$14 square. Then we upsample the output into a 224$\times$224 matrix, and apply another convolutional head to get a single channel map of size 224$\times$224.  

\subsection{The effect of scale}
In recent years it has been shown that scaling transformers allows to consistently improve prediction performance for many modalities. \cite{scalingViT} thoroughly examined the scaling laws for ViTs, and concluded that scaling both the model and the data improves representation quality. Still, they showed that the performance can be bottlenecked by the dataset size. In the second approach we used ViT-B, an 86 million parameter network. Here we attempt to use larger ViTs to see their effect. Following \cite{scalingViT}, the next scales are ViT-L/14 (303M parameters) and ViT-g/14 (1011M parameters). CLIP has been pretrained on ViT-L/14 as well, so we can initialize with its weights. 

\subsection{Training}
We followed the split of WAIR-D dataset from \cite{our-localization-BDS2023} that ensures all maps from Scenario 2 belong to the validation set. We end up with 9303 maps in the training set, 596 maps in the validation set and 100 maps in the test set. We train all models on NVIDIA DGX A100 at Yerevan State University utilizing six 40GB A100 GPUs with distributed data parallelism \cite{PyTorch-ddp}. Learning rate is fixed to $3\cdot 10^{-4}$. Batch size is selected to be the maximum that fits in the GPU memory. After training, we select the epoch with the lowest validation loss.

\section{Performance evaluation} \label{sec::perf}

\begin{table*}[t!]
\centering
\caption{The performance of our models on the test set of the WAIR-D dataset \cite{huangfu2022waird}. All models use no augmentation and use dice loss.}
\label{tab:main-results}
\begin{tabular}{@{}llcccccc@{}}
\toprule
Encoder (Params)       & Decoder (Params)          & Pretrained & Recall (\%) & Precision (\%) & IoU (\%) & Hausdorff (m) & Chamfer (m) \\
\midrule
U-Net encoder (19M)    & U-Net decoder (12M)           & \xmark & 64.4 & 26.4 & 23.3 & 94.4  & 27.9 \\
ResNet (26M)          & U-Net decoder (48M)           & \cmark & 54.9 & 33.1 & 25.5 & 113.0 & 34.3 \\ 
\midrule
ViT-B/16-CLIP+ (87M)  & UPerNet (4 layers) (40M)     & \cmark & 51.6 & 31.0 & 23.8 & 81.3  & 21.0 \\
ViT-B/16-CLIP+ (87M)  & UPerNet (all layers) (99M)   & \cmark & 65.7 & 52.0 & 41.7 & 61.6  & 15.4 \\
ViT-B/16-CLIP+ (87M)  & UPerNet (all layers) (99M)   & \xmark & 66.5 & 52.4 & 42.2 & 64.9  & 18.3 \\ \midrule
ViT-L/14-CLIP+ (305M) & UPerNet (all layers) (301M)  & \cmark & 68.4 & 50.6 & 42.0 & 57.1  & 14.7 \\
\bottomrule
\end{tabular}
\end{table*}

\begin{table*}[t!]
\centering
\caption{Experiments on U-Net-based networks.}
\label{tab:unet-additional-results}
\begin{tabular}{@{}llccccc@{}}
\toprule
Augmentation                & Loss      & Recall (\%) & Precision (\%) & IoU (\%) & Hausdorff (m) & Chamfer (m) \\
\midrule
None                        & Dice      & 64.4 & 26.4 & 23.3 & 94.4  & 27.9 \\
translation, rotation, flip & Dice      & 52.4 & 28.5 & 20.4 & 115.3 & 44.9 \\
translation, rotation, flip & Dice + CE & 35.6 & 35.7 & 18.2 & 149.7 & 78.3 \\ 
\bottomrule
\end{tabular}
\end{table*}


\subsection{Quantitative analysis}
To evaluate the predictions we binarize the outputs by applying a threshold of 0.5 and calculate the metrics defined in Section \ref{sec::problem}. The results are summarized in Table \ref{tab:main-results}.

First we note that convolutional approaches start to overfit in early epochs and the best performance on the validation set is achieved between epochs 5 and 8. The best performing U-Net-based approach reached 23.3\% IoU on the test set and 27.9m Chamfer distance. Basic image augmentation and loss function changes didn't improve the performance (Table \ref{tab:unet-additional-results}).

The version with ImageNet-pretrained ResNet achieved slightly better IoU, but Chamfer distance got worse. Note that this convolutional network has more parameters than the regular U-Net.

The initial version of the transformer-based approach that utilizes regular UPerNet attached to four layers of the ViT, did not bring improvements in terms of IoU, but had a significantly better Chamfer distance (21.0m). The loss curves showed that this model is \textit{underfitting}. This motivated us to ease the flow of information between encoder and decoder by introducing more connections. By attaching UperNet to all 12 layers of the ViT, we increased the parameter count by another 60M, and got a significant improvement to 41.7\% IoU and 15.4m Chamfer distance. 

As an ablation, we tried to train the same network without initializing ViT-B from CLIP. We got a small improvement in terms of IoU (+0.5\%), but Chamfer distance degraded by 3 meters. While this is not surprising due to the significant mismatch in data distributions of regular images used in CLIP and our inputs, it indicates that transfer learning from pretrained models on the image datasets should be seriously considered for wireless communication problems.

Finally, we moved from ViT-B/16 to ViT-L/14 and got a small improvement both in terms of IoU (+0.3\%) and Chamfer distance (-0.7m). Notably, this model started to overfit much earlier, which indicates that the current dataset size is already a bottleneck for this model size. The results are consistent with the results on ImageNet-21k from \cite{scalingViT}. While the dataset we used is computer simulated, this result indicates that it is possible to improve the reconstruction performance by generating larger and more diverse datasets and using larger transformers.

\subsection{Qualitative analysis}
\begin{figure*}
    \centering
    \includegraphics[trim={4cm 4cm 4cm 4cm},clip,width=\textwidth]{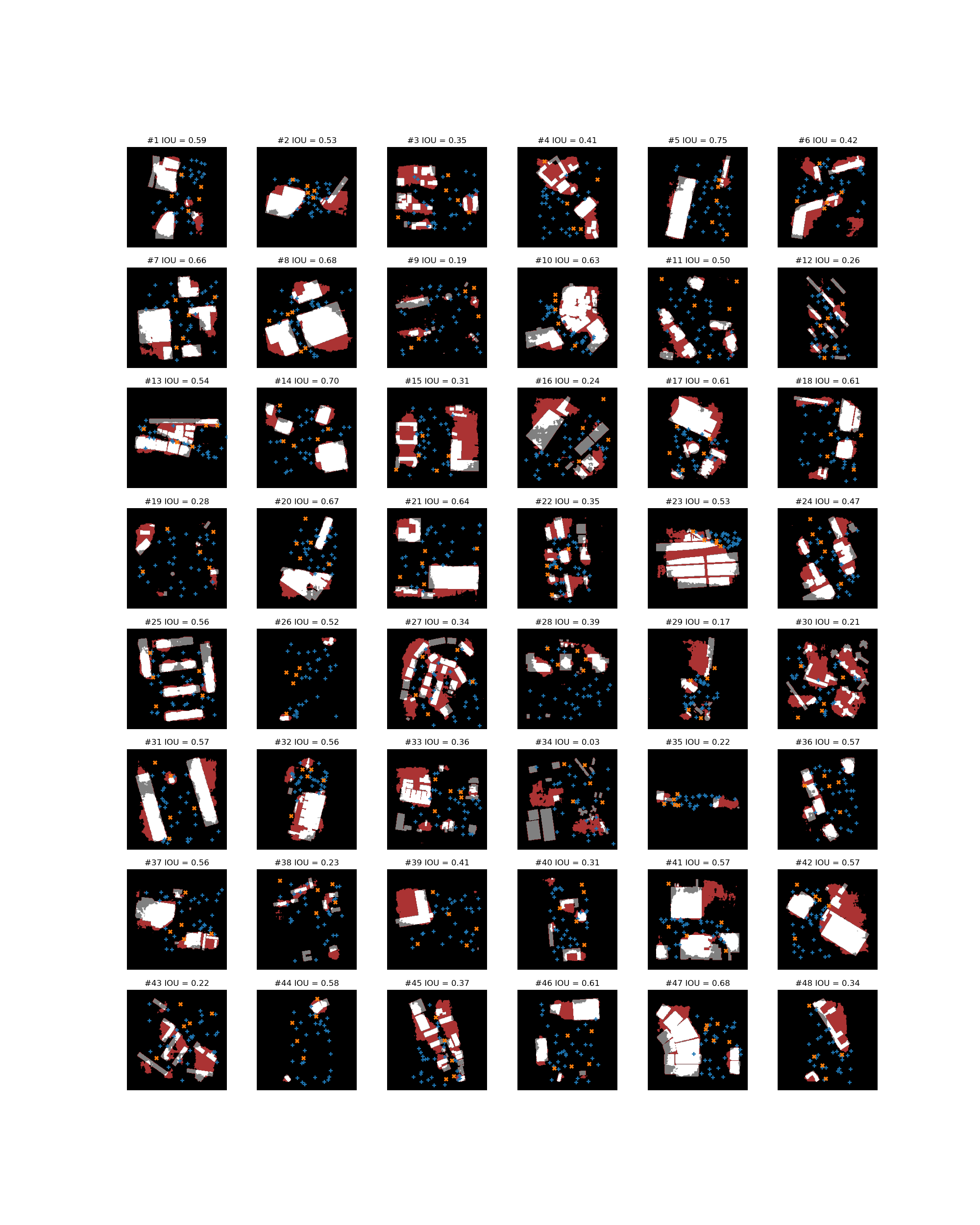}
    \caption{Outputs of the best model (CLIP-initialized ViT-L/14 with UPerNet decoder). White and grey pixels indicate buildings. Grey pixels indicate parts of the buildings not detected by our model (false negatives). Dark red pixels indicate false positives. Orange and blue crosses indicate locations of base stations and user equipments, respectively.}
    \label{fig:ViTL-results}
\end{figure*}

Figure \ref{fig:ViTL-results} shows the ground truth maps overlaid with predictions of our largest model on 49 environments from the test set. Visual examination shows the following patterns.

(1) The model accurately predicts the faces of the buildings facing a large concentration of devices. On the contrary, the other faces of the buildings behind the devices and antennas are predicted pretty much randomly. The buildings on the upper half of the maps \#1, \#33, \#41 and \#48 are quite indicative. Note that the predictions of the ``hidden'' faces of the buildings go wrong in both ways: the model can ``make'' the buildings more thick (\#33, \#48) and more narrow (\#1) compared to their original sizes. We believe this kind of errors are almost impossible to solve with better prediction models as there is a lack of information.

(2) The model struggles to predict gaps between buildings, especially if the gaps are small and there are no devices between the buildings. The model ``fills in'' the gaps and predicts larger and longer buildings. Maps \#4, \#15, \#24 and \#46 clearly demonstrate this issue. In contrast, the gaps are correctly predicted on maps \#8 and \#25 where the gaps contain several devices. Whether the available radio link information is sufficient to improve this aspect of the predictor models is an open question.

(3) Whenever a device is within line-of-sight from an antenna, the model correctly marks the entire line as an empty area. There are very few exceptions to this, e.g. in the upper-right part of the map \#25.


\section{Acknowledgement}
The work of H. Khachatrian and R. Mkrtchyan was partly supported by the RA Science Committee grant No. 22rl-052 (DISTAL). The work of T. P. Raptis was partly supported by the European Union under the Italian National Recovery and Resilience Plan (NRRP) of NextGenerationEU, partnership on ``Telecommunications of the Future'' (PE00000001 - program ``RESTART'').

\balance

\bibliographystyle{IEEEtran}
\bibliography{refs}

\end{document}